\documentclass[11pt,twoside]{article}
\usepackage{asp2004}
\usepackage{psfig}
\usepackage{epsf}
\usepackage{graphics}
\usepackage{lscape}
\def\edcomment#1{\iffalse\marginpar{\raggedright\sl#1\/}\else\relax\fi}
\reversemarginpar

\begin{document}
\title{Dynamic X-ray spectra of Chandra cataclysmic variables}
\author{A. V. Halevin}
\affil{Department of Astronomy, Odessa National University, T.G.Shevchenko park, 65014, Odessa, Ukraine
}

\begin{abstract}
In this work we have tested conception of energy dependent power spectra which can 
be applied for analysis of X-ray dynamic spectra of cataclysmic variables. Using this 
technique we can study properties of variability in different energy ranges.
\end{abstract}
\thispagestyle{plain}

\section{Introduction}

Taking  into  account rapidly growing sensitivity of X-ray detectors we can
apply to X-ray data more  complicated  techniques of time series analysis. Especially it
could  be  useful  to analyse properties of variability in different energy
ranges  to  detect different  physical  mechanisms which form behaviour of
X-ray  sources.  The idea of energy-dependent time series analysis use
binning of X-ray data both  in  time  and  energy  space (so  called ``dynamic
spectra'', \citet{hal2004}). As a result, we can calculate power spectra of light curves for
each (narrow enough) energy  bin. Resulting set of power spectra can be showed as a function
of  two  variables:  energy  and  time-scale.  Such  product  we called as
``energy-dependent power spectra'' (EDPS).

\section{TESTING OF EDPS}

To test capabilities of proposed technique we have modeled dynamic spectrum 
as a product of two variable power laws with different photo-indexes (Eq.1).

\begin{eqnarray} \nonumber
I(E,t) = PL_1(E,t)+ PL_2(E,t)\\
PL_i(E,t) = E^{\gamma_i(t)}\\   \nonumber
\gamma_i(t)=\gamma_{0i} + A_i \cdot \gamma_{0i} \cdot \sin(\frac{2\pi}{P_i}t)\\ \nonumber
\end{eqnarray}

\noindent
where $E$ is energy, $t$ is time, $PL_i(E,t)$ and
$\gamma_i(t)$ are power law functions and photo-indexes and $P_i$ are periods. 
In the model we have used 
the next values $\gamma_{01}=2.0$, $\gamma_{02}=1.0$, $A_{1,2}=5\%$, $P_1=30000$ sec 
and $P_2=10000$ sec. Resulting EDPS one can see in Fig.1.

\begin{figure}[!t]
\vspace{0cm} \hspace{2cm} 
\resizebox{9cm}{!}{\includegraphics{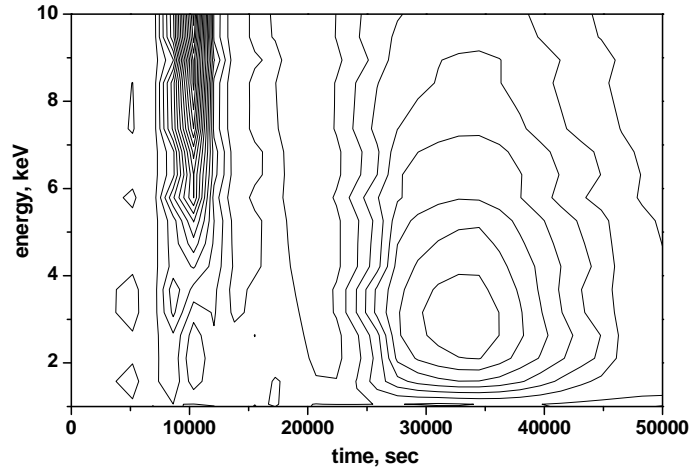}} \caption{Energy dependent 
power spectra map, calculated for variability model (1).} \label{fig1} 
\vspace{0cm} 
\end{figure}

\begin{figure}[!t]
\vspace{0cm} \hspace{2cm} 
\resizebox{9cm}{!}{\includegraphics{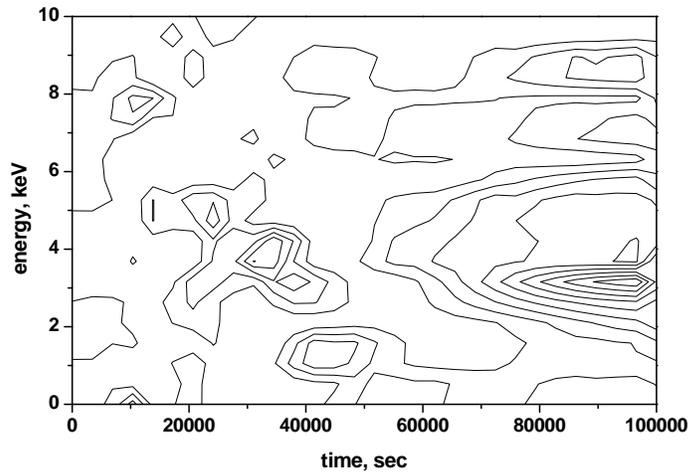}} \caption{Energy dependent 
power spectra map for Chandra observations of IP GK Per.} \label{fig2} 
\vspace{0cm} 
\end{figure}

Here the steeper power law with longer variability time-scale is 
detected on low energies. On high energies dominates variability of power law 
component with $\gamma = 1.0$.

In  Fig.2 we showed EDPS map for Chandra observations of intermediate polar
GK  Per  (archive  dataset  650).  Here quantum statistics is not enough to
analyse  spin  variability  of  GK  Per.  Detected at energies about 3 keV
variations  with  time-scale  larger  than 80 ksec corresponds to orbital
variability trend of  X-ray  flux. So we  can  conclude  that  EDPS  is perspective
technique  to analyse data of future X-ray missions with higher sensitivity
of detectors.

\end{document}